\documentclass[revtex, amsmath, amssymb,superscriptaddress, reprint]{revtex4-2}

\usepackage[utf8]{inputenc}
\usepackage{graphicx}
\usepackage{dcolumn}
\usepackage{bm}
\usepackage[utf8]{inputenc}
\usepackage[T1]{fontenc}
\usepackage{mathptmx}
\usepackage{gensymb}
\usepackage{upgreek}
\usepackage{xcolor}
\usepackage{textcomp}
\usepackage{hyperref}
\usepackage{lineno}
\usepackage{mathtools,amssymb}
\usepackage{siunitx}
\usepackage{threeparttable}
\usepackage{xcolor}
\usepackage{booktabs} 

\hypersetup{colorlinks=true,linkcolor=red,citecolor=blue,filecolor=magenta,urlcolor=magenta}

\usepackage{titlesec}
\titleformat{\section}[hang]{\small\bfseries\sffamily}{\thesection.}{0.5em}{\MakeUppercase}
\titlespacing{\section}{0pc}{1pc}{0.2pc}
\usepackage{soul}

\begin{document}

\author{Likai Yang}
\affiliation{Department of Electrical Engineering, Yale University, New Haven, CT 06511, USA}
\author{Chunzhen Li}
\affiliation{Department of Electrical Engineering, Yale University, New Haven, CT 06511, USA}
\author{Jiacheng Xie}
\affiliation{Department of Electrical Engineering, Yale University, New Haven, CT 06511, USA}
\author{Hong X. Tang}
\email{hong.tang@yale.edu}
\affiliation{Department of Electrical Engineering, Yale University, New Haven, CT 06511, USA}

\title{High-Q and Compact Fabry-Perot Microresonators on Thin-Film Lithium Niobate}

\begin{abstract}
Thin-film lithium niobate (TFLN) has played a pivotal role in the advancement of integrated photonics, by supporting a diverse range of applications including nonlinear optics, electro-optics, and piezo-optomechanics. The effective realization and enhancement of these interactions rely heavily on the implementation of high quality photonic microresonators. The pursuit of novel resonator architectures with optimized properties thus represents a central research area in TFLN photonics. In this work, we design and fabricate TFLN Fabry-Perot microresonators, by placing a straight section of waveguide between a pair of tapered photonic crystal mirrors. The resonator features a high quality factor of 6$\times$10$^5$ at 1530\,nm and a compact length of 100\,{\textmu}m. The functionality of the device is further demonstrated by integrating on-chip electrodes for high-frequency piezo-optomechanical modulation. Our device can serve as an appealing candidate for developing high-performance photonic components on the TFLN platform.  
\end{abstract}

\maketitle

\section{Introduction} 
A key focus of modern integrated photonics is to engineer interactions between optical fields and other physical systems such as microwave and acoustics. Lithium niobate, as a material with advantageous linear and nonlinear optical properties, has been extensively investigated for this purpose. Recent advancement in lithium niobate on insulator (LNOI) thin films, along with the direct etching technique for fabricating high-index-contrast optical waveguides, have further broadened the scope of its applications. Lithium niobate exhibits a wide transparency window, typically ranging from 350\,nm in the ultraviolet to 4000\,nm in the mid-infrared. Its high refractive index (n$\approx$2.2 at 1550\,nm) also enables the realization of optical waveguides on low-index substrates such as silicon dioxide, as in the case of LNOI thin films. Owing to its inherent $\chi^{(2)}$ and $\chi^{(3)}$ nonlinearity, lithium niobate plays a vital role in nonlinear optics, for purposes such as second harmonic generation \cite{wang2017second,luo2018highly}, supercontiniuum generation \cite{lu2019octave,yu2019coherent}, and Kerr frequency combs \cite{wang2019monolithic}. Efficient and broadband electro-optic modulators, which are essential components for optical telecommunication, have also been developed based on the pronounced Pockels effect of lithium niobate \cite{wang2018integrated}. Its piezoelectric nature also facilitates the study of cavity optomechanical devices \cite{jiang2019lithium}, which are valuable resources for precision sensing and quantum technologies. Furthermore, lithium niobate can serve as host materials for rare earth emitters to build lasers \cite{luo2021chip}, optical amplifiers \cite{cai2021erbium}, and single photon sources \cite{yang2023controlling}. In light of this broad spectrum of functionalities, significant attention is directed toward the development and optimization of photonic devices based on LNOI platforms.

The advancement of LNOI photonic technologies heavily relies on the implementation of high-quality microresonators, which facilitate resonance enhancement of light-matter interaction to improve device performance. In many cases, device optimization hinges on the design and realization of specific resonator types. The most commonly adopted photonic resonators on LNOI platform are the whispering-gallery-mode (WGM) resonators, including microdisks \cite{wang2014integrated}, microrings \cite{zhang2017monolithic}, and racetracks \cite{pan2020compact}. To date, WGM resonators with quality factor (Q) over a million have been realized and applied to aforementioned use cases including frequency comb generation \cite{gong2020near}, efficient microwave-to-optical conversion \cite{mckenna2020cryogenic,holzgrafe2020cavity,xu2021bidirectional}, and piezo-optomechanics \cite{jiang2016chip,shen2020high}. For certain applications, however, the inherent limitations of WGM resonators may impose challenges on achieving high-performance devices. One important constraint arises from the anisotropy of lithium niobate, which prevents a full mode overlap between optical fields and the targeted microwave or acoustic modes. For example, a preferable configuration for electro-optic modulation on LNOI is to use x-cut films and apply in-plane electric field along the crystal z-axis, thereby utilizing the largest electro-optic coefficient $r_{33}$ \cite{lenzo1966electro}. In this scenario, for microring based devices only 50\,\% of the total length contributes to the modulation. The birefringence nature of lithium niobate will also impose complexity in dispersion and phase-matching engineering of WGM resonators, which are crucial factors in nonlinear-optics devices. Moreover, the mode size of WGM resonators is limited by increasing radiation loss and sidewall scattering loss as the resonator radius is reduced.

An alternative structure that is advantageous for lifting these obstacles is the Fabry-Perot (FP) microresonator, which is formed by placing mirrors at both ends of a straight waveguide. Using this structure, the device orientation can be easily engineered to match the strongest nonlinearity; the cavity size is also widely tunable by adjusting the length of the waveguide. The bottleneck in achieving low-loss and compact on-chip FP cavity is the realization of mirrors with high reflectivity and small footprint. Previous demonstrations of FP microresonators on LNOI have adopted either Bragg gratings \cite{yu2023chip} or loop mirrors \cite{cheng2024efficient}. Both approaches, however, are suboptimal in terms of reflectivity and size. As a result, the Q factor of cavities with over 1\,mm length are limited to $10^5$. On the other hand, photonic crystal structures featuring high index contrast has been realized on LNOI for detect cavities \cite{liang2017high}, providing a pathway to high-reflectivity mirrors with minimal length.

\begin{figure}[!t]
\includegraphics[width=0.5\textwidth]{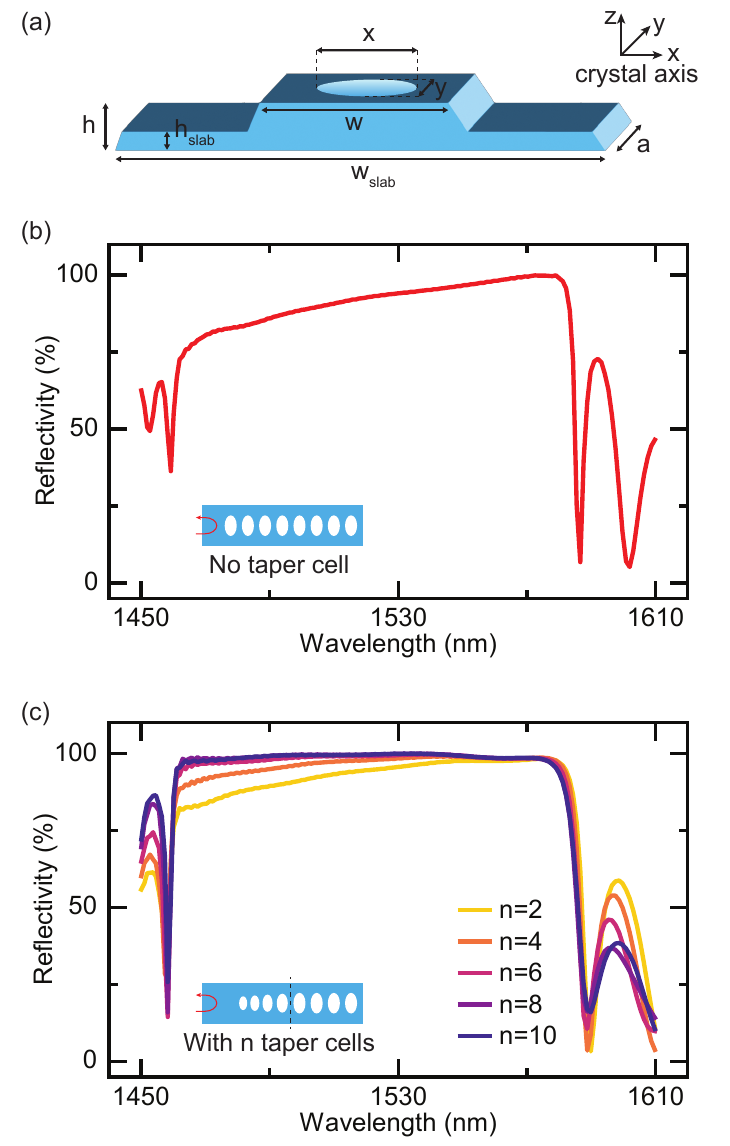}
\caption{(a) Schematic drawing of the photonic crystal unit cell, together with the direction of lithium niobate crystal axis. It consists of a suspended ridge waveguide with air holes etched through at the center. The lattice constant of the unit cell is $a=498$\,nm and the waveguide width is $w$=1.2\,{\textmu}m. All other dimensions can be found in the text. The optical mode studied here is the fundamental TE mode with electric field components mainly along the x-axis. (b) Simulated reflection spectrum of the photonic crystal, formed by cascading 30 unit cells. The reflectivity is clearly suboptimal with poor performance at shorter wavelength. (c) The reflectivity is significantly improved by linearly tapering down the lattice constant and hole size at the cavity side. The number of taper cells $n$ is varied to show the improvement. No apparent change in reflection is seen when $n$>8, we thus choose $n$=10 for the FP resonator device.}
\label{fig1}
\end{figure}

In this work, we design and fabricate high-Q LNOI FP microresonators based on photonic crystal mirrors. With a finite-difference time-domain (FDTD) simulation, We show that mirrors with near-unity reflectivity can be realized using a tapered photonic crystal structure. The fabricated 100\,{\textmu}m-long FP cavity, which has a free spectral range (FSR) of 4.8\,nm, exhibits a high loaded Q of 5.7$\times$10$^5$ at 1530\,nm. To explore the application space of our device, we further integrate on-chip electrodes to piezoelectrically excite the mechanical modes supported by the suspended waveguide. The optomechanical modulation spectrum is then measured, showcasing the presence of high-frequency thickness mode up to 19\,GHz. Our results pave the way for the development of FP microresonator based LNOI devices and are attractive for applications including electro-optics and piezo-optomechanics.

\section{Device design and fabrication}

Our device is designed on 300\,nm-thick z-cut lithium niobate thin films, with 2\,{\textmu}m-thick silicon dioxide on silicon as substrates. The photonic crystal mirrors, whose unit cell structure is sketched in Fig.~\ref{fig1}a, are realized by patterning through holes on a half-etch ridge waveguide. To get a better mode confinement, we suspend the waveguide by removing the silicon dioxide beneath it. The dimensions of the unit cell are designed to have a bandgap at telecom wavelength near 1550\,nm. To be specific, the lattice constant is set to $a$=498\,nm. The width of the waveguide and the slab are $w$=1.2\,{\textmu}m and $w_{\mathrm{slab}}$=3\,{\textmu}m, respectively. The hole size is $x\times y$=600$\times$350\,nm and the thickness of the slab is $h_{\mathrm{slab}}$=120\,nm. The waveguide exhibits a sidewall angle of 60$^{\circ}$ from the etching process. We note that the optical mode we consider here is the fundamental transverse-electric (TE) mode with the electrical field along the crystal x-axis.

\begin{figure}[!b]
\includegraphics[width=0.5\textwidth]{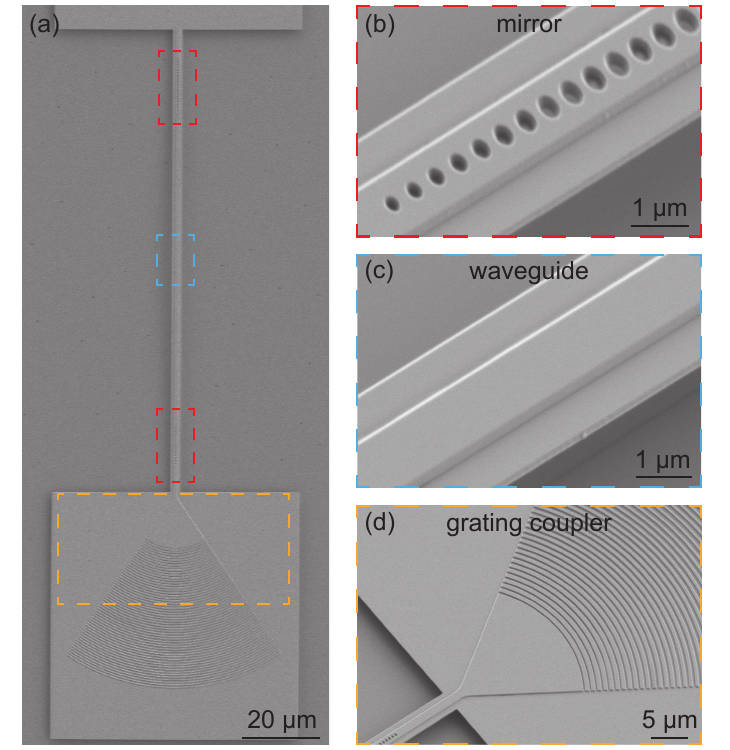}
\caption{(a) SEM image of the device. The device consists of a straight waveguide placed in between two photonic crystal mirrors. The waveguide is suspended and supported by two pads at each ends. A grating coupler is placed at one side after a short section of bus waveguide. The number of unit cells is reduced for the mirror on the coupling side, allowing access to the resonance via reflection measurement. The red, blue, and orange boxes correspond to the zoom-in view of components in (b)-(d). (b) Zoom-in view of the photonic crystal mirror. A taper structure is implemented on the cavity side. (c) Zoom-in view of the waveguide. (c) Zoom-in view of the grating coupler. It has an apodized structure and exhibits single-pass coupling efficiency of 35\,\% at center wavelength 1530\,nm.}
\label{fig2}
\end{figure}

\begin{figure*}[!t]
\includegraphics[width=1\textwidth]{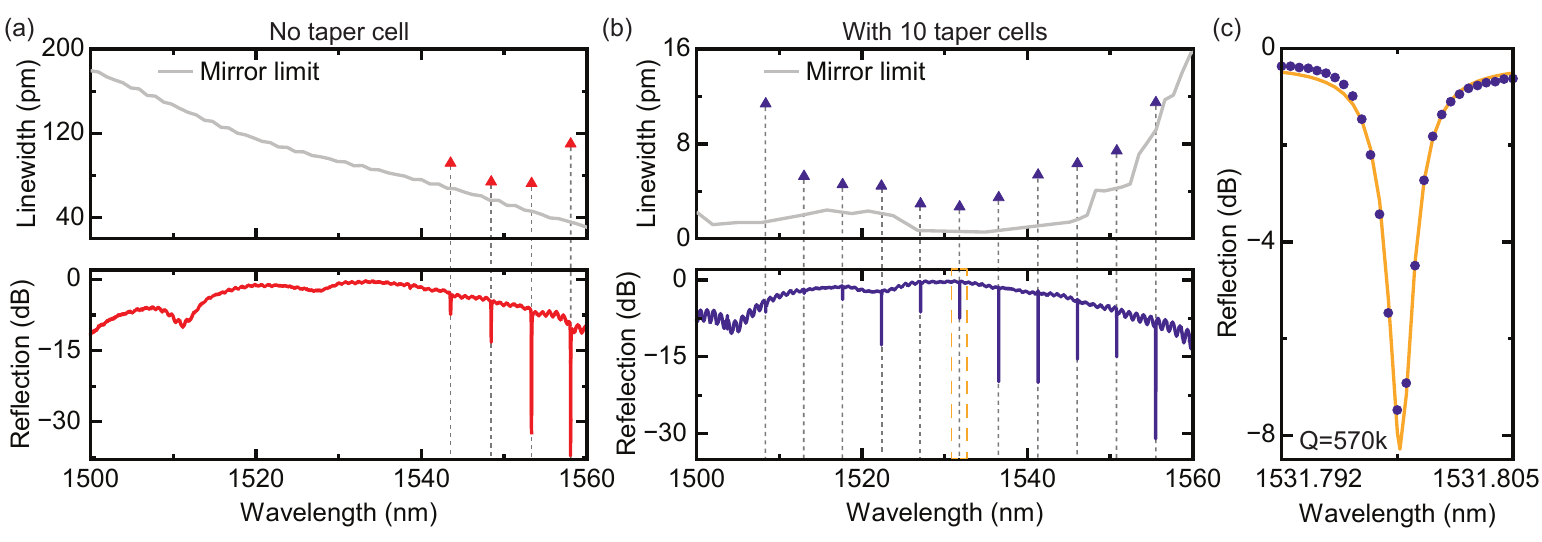}
\caption{(a) Measurement results for the FP cavity without the implementation of tapered mirrors. The bottom plot is the reflection spectrum; the top plot shows the fitted linewidth together with the lower limit calculated from mirror reflectivity simulation. The resonances exhibit low Q factor and are only seen at longer wavelength, which aligns with the prediction from simulation. (b) Measurement results for the device with tapered mirrors. The resonator Q is significantly improved compared to the case in (a). The trend of resonance linewidth agrees qualitatively with the simulated lower limit. The dashed orange box indicates the highest-Q resonance plotted in (c). (c) The resonance at 1531.8\,nm, with a loaded Q of 5.7$\times$10$^5$ and 8.3\,dB extinction ratio. The orange line is a Lorentzian fitting.}
\label{fig3}
\end{figure*}

To evaluate the performance of the photonic crystal as a mirror, its reflectivity is extracted by a FDTD simulation (Ansys Lumerical) of 30 cascaded unit cells. The results are shown in Fig.~\ref{fig1}b. While a bandgap is present from 1460\,nm to 1580\,nm, the reflectivity is low at shorter wavelength and only reaches unity at the other edge of the bandgap. This deficiency can be attributed to the index mismatch at the intersection between the waveguide and the mirror, where the abrupt change will result in excessive scattering loss. Here, we propose to mitigate the issue by introducing the taper cells. i.e. The lattice constant and the hole dimension are linearly tapered down to $450$\,nm and 400$\times$233\,nm, respectively, at the cavity side over a group of $n$ unit cells. They are then combined with the constant unit cells. The simulated reflection spectra after modification are plotted in Fig.~\ref{fig1}c, with a varying number of taper cells $n$. The reflectivity within the bandgap becomes higher and flatter as $n$ increases, clearly indicating the advantages of the tapering. Note that reflection spectrum does not change noticeably when $n$ is large enough. We thus choose $n$=10 taper cells in the FP cavity implementation. Detailed analysis of photonic bandgap structure can be found in previous works that studied similar geometry \cite{yang2024quantum,li2020lithium}.

The fabrication of our device is done by a combined process of ebeam lithography (EBL) and reactive ion etching (RIE). The ridge waveguide is first patterned using hydrogen silsesquioxane (HSQ) as a mask, followed by RIE of lithium niobate with argon (Ar). The air holes are then created following the same process, but with lithium niobate fully etched instead of half etched. After dry etching, the device is cleaned in RCA-1 solution to remove the redeposition during the RIE process. Finally, the cavity is fully suspended by dipping into buffered oxide etch (BOE) and subsequently dried in a critical point dryer.

The SEM image of the fabricated device is shown in Fig.~\ref{fig2}, with a zoom-in view of the photonic crystal mirrors, the straight waveguide, and the grating coupler for fiber-to-chip interface. The device is a fully suspended FP microresonator supported by two pads at each ends. The length of the cavity is designed to be 100\,{\textmu}m but can be tuned freely to achieve a desired FSR. The number of unit cell for the bottom mirror is reduced to slightly increase the transmittance. The cavity modes can then couple to the bus waveguide and be accessed by measuring the reflection. The grating coupler placed at the end of the bus waveguide is designed with an apodized structure \cite{lomonte2021efficient}. It has a single-pass coupling efficiency of 35\,\% at 1530\,nm and a 3-dB bandwidth of 35\,nm.

\begin{figure*}[!t]
\includegraphics[width=1\textwidth]{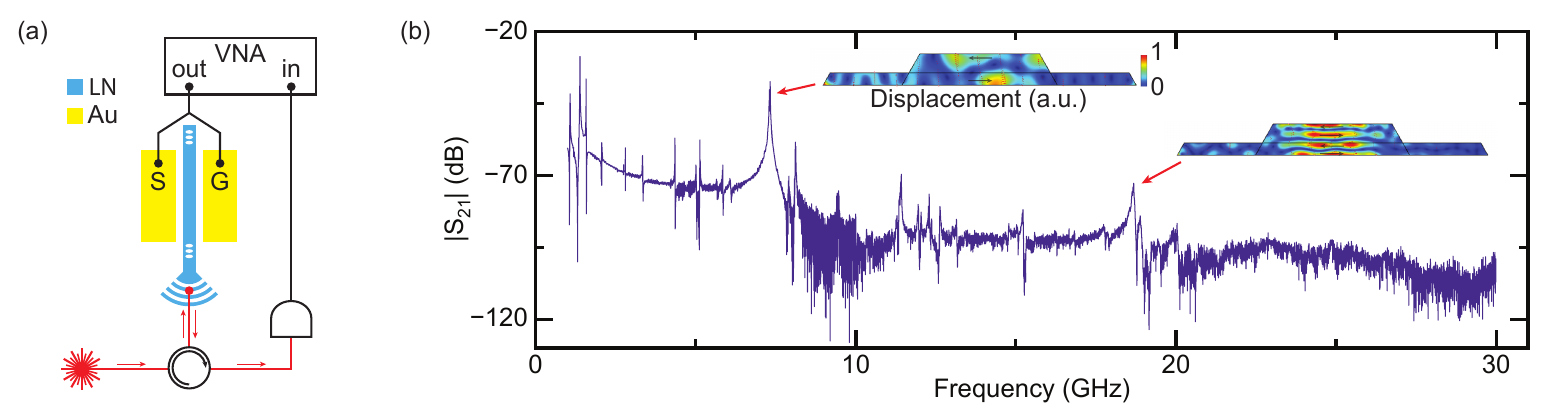}
\caption{(a) Experimental setup for measuring piezo-optomechanical modulation. The vector network analyzer (VNA) applies RF output to the on-chip gold (Au) electrodes via an RF probe. It then excites the mechanical modes supported by the suspended waveguide via piezoelectric effect. By tuning the laser into optical resonance, the light will experience optomechanical modulation, which is measured via the beating signal after the high-speed photodetector. (b) The piezo-optomechanical modulation spectrum measured up to 30\,GHz. The prominent peaks are the first-order and third-order thickness modes at 7.3\,GHz and 18.7\,GHz, respectively, which are most strongly coupled to the RF excitation. Their normalized displacement in the waveguide cross section is shown and are mostly along the crystal x-axis.}
\label{fig4}
\end{figure*}

\section{Results}
The resonance spectra of the fabricated devices are measured by sending a tunable laser and collecting the reflection via an optical circulator. For direct comparison, devices without taper cells are also fabricated and measured, with the results shown in Fig.~\ref{fig3}(a). We also plot the Lorentzian-fitted linewidth of each resonance together with the lower limit from the simulated mirror reflectivity ($R$). This is done by estimating the finesse ($\mathcal{F}$) of the FP cavity using
\begin{equation}
    \mathcal{F}=\frac{\pi\sqrt{R}}{1-R}.
    \label{eq1}
\end{equation}
The linewidth is then calculated as $\delta\lambda=\lambda_{\mathrm{FSR}}/\mathcal{F}$, where $\lambda_{\mathrm{FSR}}=4.8$\,nm is the free spectral range extracted from the measured spectrum. As seen from the measurement, cavity resonances are only visible at longer wavelength. This is consistent with the prediction from the simulation, in which the mirror reflectivity increases with wavelength. Still, the resonances exhibit linewidth over 70\,pm, corresponding to a low Q factor below 2.2$\times$10$^4$. 

Measurement results from the device implementing the taper structure are plotted in Fig.~\ref{fig3}b. With this design, an over 20-fold improvement in Q factor is achieved. The highest-Q resonance at 1531.8\,nm is highlighted in Fig.~\ref{fig3}c. It has a loaded Q of 5.7$\times$10$^5$, i.e. 2.7\,pm linewidth, with an extinction ratio of 8.3\,dB. While the spectral flatness of the mirror reflectivity is significantly improved by applying the taper, it still shows some wavelength dependency as seen in the upper plot of Fig.~\ref{fig3}b. The trend in measured linewidth agrees qualitatively with the simulated limit imposed by the reflectivity. The wavelength at which the reflectivity maximizes can be tuned by the parameters of the taper, i.e. the end lattice constant and hole size.

To demonstrate the potential applications of the FP microresonator, we study its functionality as a piezo-optomechanical modulator. This is achieved by patterning 50\,nm-thick gold electrodes along the waveguide with a PMMA liftoff process. The modulation spectrum is then measured via the setup illustrated in Fig.~\ref{fig4}a. The signal from a vector network analyzer (VNA) is first sent to the electrodes through an RF probe, so as to excite the mechanical modes supported by the waveguide via piezoelectric effect. Meanwhile, the laser is tuned into the cavity resonance, and the modulated light from the device is collected by a high-speed photodetector to obtain the beating signal. The piezo-optomechanical modulation spectrum is thus the $S_{21}$ measured by the VNA. The results are plotted in Fig.~\ref{fig4}b. The spectrum reveals multiple peaks corresponding to the mechanical modes supported by the waveguide. Two dominant peaks are the first-order and third-order thickness modes \cite{yang202010,xie2023sub} at 7.3\,GHz and 18.7\,GHz, respectively. Their simulated cross sectional displacement is shown in the inset of Fig.~\ref{fig4}b, with displacement mostly in the horizontal direction along the crystal x-axis. They are most strongly coupled to the electric field via the largest piezoelectric coefficient $e_{15}$ \cite{bouchy2022characterization}. Limited by the bandwidth of the setup, only up to third-order thickness mode is visible in the spectrum, which can be improved to investigate optomechanical coupling to the higher-order modes.

\section{Conclusion and discussion}

\begin{table} [!b]
\caption{Performance comparison of LNOI microresonators near 1550\,nm.}
\begin{threeparttable}
\begin{tabular}{ccccc}
 & Type 	& Q 	& FSR (nm) & Finesse \\ \midrule
This work & FP cavity & 5.7$\times$10$^5$ & 4.8 & 1800 \\
2023 \cite{yu2023chip} & FP cavity & 1.6$\times$10$^5$ & 0.063 & 6.6 \\
2024 \cite{cheng2024efficient} & FP cavity & 1.1$\times$10$^5$ & 0.52 & 37 \\
2025 \cite{qi2025low} & FP cavity & 1.9$\times$10$^6$ & 0.15 & 180 \\
2022 \cite{shams2022reduced} & microring & 5$\times$10$^6$ (Q$_i$)\tnote{$\dagger$} & 1.2 & <3800 \\
2023 \cite{yan2023efficient} & microdisk & 1.9$\times$10$^5$ & 8.0 & 980 \\
2024 \cite{cheng2024frequency} & racetrack & 3.2$\times$10$^6$ & 0.24 & 490 \\

\end{tabular}
\begin{tablenotes}\footnotesize
\item[$\dagger$] Only intrinsic Q (Q$_i$) is reported
\end{tablenotes}
\end{threeparttable}
\label{tab1}
\end{table}

In conclusion, we have developed a free-standing FP microresonator at telecom wavelength around 1550\,nm on LNOI thin films. The cavity is realized by implementing tapered photonic crystal mirrors at both ends of a straight waveguide. It exhibits a high Q factor of 5.7$\times$10$^5$ and a compact length of 100\,{\textmu}m, resulting in a 4.8\,nm FSR. The performance of the device can be well-represented by the simulation of mirror reflectivity. By integrating on-chip electrodes along the cavity, we explore the potential of our device in optomechanical applications. Piezo-optomechanical modulation spectrum is measured and high-frequency mechanical modes up to 18.7\,GHz are identified, establishing the capacity of our device in studying high-frequency cavity optomechanics.

Tab.~\ref{tab1} summarizes the performance metrics of several state-of-the-art LNOI microresonators operating at this wavelength. A convenient figure-of-merit for comparison is the cavity finesse, which captures the trade-off between Q factor and cavity size. By improving the mirror reflectivity while reducing its size, the finesse of our device is over an order-of-magnitude higher than previously demonstrated FP microresonators and comparable to those reported in recent WGM resonators.

Our device can be envisioned for a range of applications including electro-optics and optomechanics. Here, we briefly describe how the improvement in mode overlap and finesse can be connected to the performance metrics of these applications. In cavity-based electro-optic (optomechanical) transducers, the conversion efficiency between microwave (acoustic) mode and optical mode is determined by the cooperativity \cite{jiang2019lithium,xu2021bidirectional}
\begin{equation}
    C=4n_pg_0^2/\kappa_o\kappa_\mu.
\end{equation}
Here, $n_p$ is the pump photon number and $g_0$ is the vacuum electro-optic (optomechanical) coupling rate. $\kappa_o=\omega_o/Q$ is the linewidth of the optical mode and $\kappa_\mu$ is the linewidth of the microwave (acoustic) mode. The conversion efficiency is only maximized when $C$ reaches unity. On the other hand, the vacuum coupling rate $g_0\propto \Gamma/\sqrt{V}$ \cite{lambert2020coherent}. The mode overlap factor $\Gamma$ is defined as the normalized tensor products between the optical field, microwave (acoustic) mode, and the coupling tensor. $V$ is the mode volume of the optical mode. Considering that the finesse of the optical mode $\mathcal{F}\propto Q/V$, we have $C\propto \Gamma^2\mathcal{F}$. Compared with WGM resonators, a larger mode overlap can be achieved in our device with a straight waveguide, thanks to better adaption to the material anisotropy. Together with the realization of high finesse, these improvements will directly contribute to the increase of device efficiency.

For practical implementation of our device, electro-optic modulators can easily be realized by designing the cavity on x-cut LNOI and applying in-plane electric field along the crystal z-axis. Two coupled optical modes, which are favorable for efficient microwave-to-optical transduction in triply-resonant scheme, can be achieved by implementing back-to-back FP cavities with a partially reflective mirror in the middle. Higher-order photonic bandgap can also be explored for cavity modes at visible wavelength, which are attractive for nonlinear optics applications. The reflectivity and spectral flatness of the mirror can be further improved by optimizing the taper geometry beyond the linear design, via more careful Bloch mode engineering \cite{sauvan2005modal}. In light of these characteristics, our cavity architecture broadens the design space of TFLN photonics and can serve as valuable complement to the widely used WGM resonators for certain applications.

\section*{acknowledgment}
The authors would like to thank Dr. Yong Sun, Dr. Lauren McCabe, Kelly Woods, and Dr. Michael Rooks for their assistance provided in the device fabrication. The fabrication of the devices was done at the Yale School of Engineering \& Applied Science (SEAS) Cleanroom and the Yale Institute for Nanoscience and Quantum Engineering (YINQE).

\section*{funding}
This project is supported by the Air Force Office of Scientific Research (AFOSR MURI FA9550-23-1-0338) and in part by the Defense Advanced Research Projects Agency (DARPA OPTIM HR00112320023). The part of the research that involves lithium niobate thin film preparation is supported by the US Department of Energy Co-design Center for Quantum Advantage (C2QA) under Contract No. DE-SC0012704.

\bibliographystyle{ieeetr}
\bibliography{References}

\begin{thebibliography}{10}

\bibitem{wang2017second}
C.~Wang, X.~Xiong, N.~Andrade, V.~Venkataraman, X.-F. Ren, G.-C. Guo, and M.~Lon{\v{c}}ar, ``Second harmonic generation in nano-structured thin-film lithium niobate waveguides,'' {\em Optics express}, vol.~25, no.~6, pp.~6963--6973, 2017.

\bibitem{luo2018highly}
R.~Luo, Y.~He, H.~Liang, M.~Li, and Q.~Lin, ``Highly tunable efficient second-harmonic generation in a lithium niobate nanophotonic waveguide,'' {\em Optica}, vol.~5, no.~8, pp.~1006--1011, 2018.

\bibitem{lu2019octave}
J.~Lu, J.~B. Surya, X.~Liu, Y.~Xu, and H.~X. Tang, ``Octave-spanning supercontinuum generation in nanoscale lithium niobate waveguides,'' {\em Optics letters}, vol.~44, no.~6, pp.~1492--1495, 2019.

\bibitem{yu2019coherent}
M.~Yu, B.~Desiatov, Y.~Okawachi, A.~L. Gaeta, and M.~Lon{\v{c}}ar, ``Coherent two-octave-spanning supercontinuum generation in lithium-niobate waveguides,'' {\em Optics letters}, vol.~44, no.~5, pp.~1222--1225, 2019.

\bibitem{wang2019monolithic}
C.~Wang, M.~Zhang, M.~Yu, R.~Zhu, H.~Hu, and M.~Loncar, ``Monolithic lithium niobate photonic circuits for kerr frequency comb generation and modulation,'' {\em Nature communications}, vol.~10, no.~1, p.~978, 2019.

\bibitem{wang2018integrated}
C.~Wang, M.~Zhang, X.~Chen, M.~Bertrand, A.~Shams-Ansari, S.~Chandrasekhar, P.~Winzer, and M.~Lon{\v{c}}ar, ``Integrated lithium niobate electro-optic modulators operating at cmos-compatible voltages,'' {\em Nature}, vol.~562, no.~7725, pp.~101--104, 2018.

\bibitem{jiang2019lithium}
W.~Jiang, R.~N. Patel, F.~M. Mayor, T.~P. McKenna, P.~Arrangoiz-Arriola, C.~J. Sarabalis, J.~D. Witmer, R.~Van~Laer, and A.~H. Safavi-Naeini, ``Lithium niobate piezo-optomechanical crystals,'' {\em Optica}, vol.~6, no.~7, pp.~845--853, 2019.

\bibitem{luo2021chip}
Q.~Luo, C.~Yang, R.~Zhang, Z.~Hao, D.~Zheng, H.~Liu, X.~Yu, F.~Gao, F.~Bo, Y.~Kong, {\em et~al.}, ``On-chip erbium-doped lithium niobate microring lasers,'' {\em Optics Letters}, vol.~46, no.~13, pp.~3275--3278, 2021.

\bibitem{cai2021erbium}
M.~Cai, K.~Wu, J.~Xiang, Z.~Xiao, T.~Li, C.~Li, and J.~Chen, ``Erbium-doped lithium niobate thin film waveguide amplifier with 16 db internal net gain,'' {\em IEEE Journal of Selected Topics in Quantum Electronics}, vol.~28, no.~3: Hybrid Integration for Silicon Photonics, pp.~1--8, 2021.

\bibitem{yang2023controlling}
L.~Yang, S.~Wang, M.~Shen, J.~Xie, and H.~X. Tang, ``Controlling single rare earth ion emission in an electro-optical nanocavity,'' {\em Nature Communications}, vol.~14, no.~1, p.~1718, 2023.

\bibitem{wang2014integrated}
C.~Wang, M.~J. Burek, Z.~Lin, H.~A. Atikian, V.~Venkataraman, I.-C. Huang, P.~Stark, and M.~Lon{\v{c}}ar, ``Integrated high quality factor lithium niobate microdisk resonators,'' {\em Optics express}, vol.~22, no.~25, pp.~30924--30933, 2014.

\bibitem{zhang2017monolithic}
M.~Zhang, C.~Wang, R.~Cheng, A.~Shams-Ansari, and M.~Lon{\v{c}}ar, ``Monolithic ultra-high-q lithium niobate microring resonator,'' {\em Optica}, vol.~4, no.~12, pp.~1536--1537, 2017.

\bibitem{pan2020compact}
B.~Pan, Y.~Tan, P.~Chen, L.~Liu, Y.~Shi, and D.~Dai, ``Compact racetrack resonator on linbo 3,'' {\em Journal of Lightwave Technology}, vol.~39, no.~6, pp.~1770--1776, 2020.

\bibitem{gong2020near}
Z.~Gong, X.~Liu, Y.~Xu, and H.~X. Tang, ``Near-octave lithium niobate soliton microcomb,'' {\em Optica}, vol.~7, no.~10, pp.~1275--1278, 2020.

\bibitem{mckenna2020cryogenic}
T.~P. McKenna, J.~D. Witmer, R.~N. Patel, W.~Jiang, R.~Van~Laer, P.~Arrangoiz-Arriola, E.~A. Wollack, J.~F. Herrmann, and A.~H. Safavi-Naeini, ``Cryogenic microwave-to-optical conversion using a triply resonant lithium-niobate-on-sapphire transducer,'' {\em Optica}, vol.~7, no.~12, pp.~1737--1745, 2020.

\bibitem{holzgrafe2020cavity}
J.~Holzgrafe, N.~Sinclair, D.~Zhu, A.~Shams-Ansari, M.~Colangelo, Y.~Hu, M.~Zhang, K.~K. Berggren, and M.~Lon{\v{c}}ar, ``Cavity electro-optics in thin-film lithium niobate for efficient microwave-to-optical transduction,'' {\em Optica}, vol.~7, no.~12, pp.~1714--1720, 2020.

\bibitem{xu2021bidirectional}
Y.~Xu, A.~A. Sayem, L.~Fan, C.-L. Zou, S.~Wang, R.~Cheng, W.~Fu, L.~Yang, M.~Xu, and H.~X. Tang, ``Bidirectional interconversion of microwave and light with thin-film lithium niobate,'' {\em Nature communications}, vol.~12, no.~1, p.~4453, 2021.

\bibitem{jiang2016chip}
W.~C. Jiang and Q.~Lin, ``Chip-scale cavity optomechanics in lithium niobate,'' {\em Scientific reports}, vol.~6, no.~1, p.~36920, 2016.

\bibitem{shen2020high}
M.~Shen, J.~Xie, C.-L. Zou, Y.~Xu, W.~Fu, and H.~X. Tang, ``High frequency lithium niobate film-thickness-mode optomechanical resonator,'' {\em Applied Physics Letters}, vol.~117, no.~13, 2020.

\bibitem{lenzo1966electro}
P.~Lenzo, E.~Spencer, and K.~Nassau, ``Electro-optic coefficients in single-domain ferroelectric lithium niobate,'' {\em Journal of the optical society of America}, vol.~56, no.~5, pp.~633--635, 1966.

\bibitem{yu2023chip}
S.~Yu, Z.~Fang, Z.~Wang, Y.~Zhou, Q.~Huang, J.~Liu, R.~Wu, H.~Zhang, M.~Wang, and Y.~Cheng, ``On-chip single-mode thin-film lithium niobate fabry--perot resonator laser based on sagnac loop reflectors,'' {\em Optics Letters}, vol.~48, no.~10, pp.~2660--2663, 2023.

\bibitem{cheng2024efficient}
J.~Cheng, D.~Gao, J.~Dong, and X.~Zhang, ``Efficient second harmonic generation in a high-q fabry-perot microresonator on x-cut thin film lithium niobate,'' {\em Optics Express}, vol.~32, no.~7, pp.~12118--12126, 2024.

\bibitem{liang2017high}
H.~Liang, R.~Luo, Y.~He, H.~Jiang, and Q.~Lin, ``High-quality lithium niobate photonic crystal nanocavities,'' {\em Optica}, vol.~4, no.~10, pp.~1251--1258, 2017.

\bibitem{yang2024quantum}
L.~Yang, {\em Quantum Photonic Devices With Rare Earth Incorporated Thin-Film Lithium Niobate}.
\newblock PhD thesis, Yale University, 2024.

\bibitem{li2020lithium}
M.~Li, J.~Ling, Y.~He, U.~A. Javid, S.~Xue, and Q.~Lin, ``Lithium niobate photonic-crystal electro-optic modulator,'' {\em Nature communications}, vol.~11, no.~1, p.~4123, 2020.

\bibitem{lomonte2021efficient}
E.~Lomonte, F.~Lenzini, and W.~H. Pernice, ``Efficient self-imaging grating couplers on a lithium-niobate-on-insulator platform at near-visible and telecom wavelengths,'' {\em Optics Express}, vol.~29, no.~13, pp.~20205--20216, 2021.

\bibitem{yang202010}
Y.~Yang, R.~Lu, L.~Gao, and S.~Gong, ``10--60-ghz electromechanical resonators using thin-film lithium niobate,'' {\em IEEE Transactions on Microwave Theory and Techniques}, vol.~68, no.~12, pp.~5211--5220, 2020.

\bibitem{xie2023sub}
J.~Xie, M.~Shen, Y.~Xu, W.~Fu, L.~Yang, and H.~X. Tang, ``Sub-terahertz electromechanics,'' {\em Nature Electronics}, vol.~6, no.~4, pp.~301--306, 2023.

\bibitem{bouchy2022characterization}
S.~Bouchy, R.~J. Zednik, and P.~B{\'e}langer, ``Characterization of the elastic, piezoelectric, and dielectric properties of lithium niobate from 25° c to 900° c using electrochemical impedance spectroscopy resonance method,'' {\em Materials}, vol.~15, no.~13, p.~4716, 2022.

\bibitem{qi2025low}
L.~Qi, A.~Khalatpour, J.~F. Herrmann, T.~Park, D.~Dean, S.~Robison, A.~Hwang, H.~Stokowski, D.~Serkland, M.~M. Fejer, {\em et~al.}, ``Low-loss, highly tunable sagnac loop reflectors and fabry--p{\'e}rot cavities on thin-film lithium niobate,'' {\em Optics Letters}, vol.~50, no.~16, pp.~5173--5176, 2025.

\bibitem{shams2022reduced}
A.~Shams-Ansari, G.~Huang, L.~He, Z.~Li, J.~Holzgrafe, M.~Jankowski, M.~Churaev, P.~Kharel, R.~Cheng, D.~Zhu, {\em et~al.}, ``Reduced material loss in thin-film lithium niobate waveguides,'' {\em Apl Photonics}, vol.~7, no.~8, 2022.

\bibitem{yan2023efficient}
C.~Yan, S.~Wang, S.~Zhao, Y.~Huang, H.~Zhou, G.~Deng, S.~Wang, and S.~Zhou, ``Efficient and temperature-tunable second-harmonic generation in a thin film lithium niobate on insulator microdisk,'' {\em Applied Physics Letters}, vol.~122, no.~10, 2023.

\bibitem{cheng2024frequency}
R.~Cheng, M.~Yu, A.~Shams-Ansari, Y.~Hu, C.~Reimer, M.~Zhang, and M.~Lon{\v{c}}ar, ``Frequency comb generation via synchronous pumped $\chi$ (3) resonator on thin-film lithium niobate,'' {\em Nature Communications}, vol.~15, no.~1, p.~3921, 2024.

\bibitem{lambert2020coherent}
N.~J. Lambert, A.~Rueda, F.~Sedlmeir, and H.~G. Schwefel, ``Coherent conversion between microwave and optical photons—an overview of physical implementations,'' {\em Advanced Quantum Technologies}, vol.~3, no.~1, p.~1900077, 2020.

\bibitem{sauvan2005modal}
C.~Sauvan, G.~Lecamp, P.~Lalanne, and J.-P. Hugonin, ``Modal-reflectivity enhancement by geometry tuning in photonic crystal microcavities,'' {\em Optics express}, vol.~13, no.~1, pp.~245--255, 2005.

\end{thebibliography}

\end{document}